# The Sociotype, a New Conceptual Construct on Human Social Networks: Application in Mental Health and Quality of Life

del Moral R[1*], Navarro J[1], López-del Hoyo Y[2], Gómez-Quintero JD[3], Garcia-Campayo J[2], Marijuán PC[1*]

[1]Bioinformation and Systems Biology Group, Aragon Health Sciences Institute (IACS-IIS Aragon)
[2]Research on Mental Health in Primary Care Group, Aragon Health Sciences Institute (IACS-IIS Aragon)
[3]GESES Group, Social Science and Labor Faculty, University of Zaragoza

*Correspondence concerning this article should be addressed to:
Raquel del Moral; Pedro C. Marijuán
CIBA Building. Avda. San Juan Bosco 13, 50009, Zaragoza (Spain)
Phone: (+34) 976 714476
Fax: (+34) 976 715554
E-mail: rdelmoral.iacs@aragon.es
pcmarijuan.iacs@aragon.es

**Abstract**

*The present work discusses the pertinence of a "sociotype" construct, both theoretically and empirically oriented. The term, based on the conceptual chain genotype-phenotype-sociotype, suggests an evolutionary preference in the human species for some determined averages of social relationships. This core pattern or "sociotype" has been explored herein for the networking relationships of young people--165 university students filling in a 20-items questionnaire on their social interactions. In spite that this is a preliminary study, interesting results have been obtained on gender conversation time, mental health, sociability level, and satisfaction with personal relationships. This sociotype hypothesis could be a timely enterprise for mental health and quality of life policies.*

# 1. INTRODUCTION

Sociality is an obvious trait of the human species --as Aristotle put in *The Politics* "man is by nature a political animal" (Fowler and Schreiber 2008). Most of the evolutionary and cultural novelties of our past refer to essential aspects of sociality --e.g. origins of language, emotional communication, group behavior, morals and ethics, religious and legal codes, political institutions, and so on. So fluid and culturally variable are the emerging structures of human sociality that, apparently, they defy any precise classification or numeric specification. Traditionally, a number of schools of thought have followed culturally-oriented approaches to the 'open ended' phenomenon of sociality (Derridá 1976; Lévi-Strauss 1990), while some others have emphasized views closer to biological determinism (Lorenz 1965; Wilson 1977). It is the old conflict between biological and political views, the "nature" versus "nurture" unfortunate dichotomy. Scientific discussions have been compounded by the many different fields of study involved --anthropological, neurobiological, ethological, psychological, social and political, economics, network science, etc. More recently, however, some anthropological and social science studies have achieved an interesting convergence between, say, biology and politics about fundamentals of human sociality (Chapais 2008, 2011). Hypothesis such as the "social brain" have also contributed to advance a new bond-centered approach on the evolutionary emergence of human sociality (Dunbar 2004; Fowler and Schreiber 2008). The presence of a series of significant regularities in the size and structures of social groups, notwithstanding their remarkable variability, suggests the plausibility of a "deep structure" of social bonding for the human species (Chapais, 2011; Hill et al, 2011). There seems to be an average of social networking, with very ample upper and lower limits, concerning the number and classes of bonding relationships that an individual is able to maintain meaningfully (Dunbar 2004; Dunbar and Shultz 2007). The finding of networking regularities such as the famous "Dunbar's number" (150-200 individual acquaintances) makes a lot of evolutionary and anthropological sense. This preliminary study explores these findings as well as other issues related to interpersonal communication and social bonding, integrating them within the whole framework of the sociotype hypothesis. In the background, two basic questions emerge: How much do we talk? With whom?

*1.1. The social brain hypothesis*

The social brain hypothesis has posited that, in primate societies, selection has favored larger brains and more complex cognitive capabilities as a mean to cope with the challenges of social life (Silk 2007). In primate societies, a tight correlation has been observed between the size of social groups and the neocortex relative proportion (roughly, "brain size"). Actually, the idea of relating brain size with the demands of communication in social life was already hinted by C. Darwin in "The Descent of Man" (1871). More than a century later, J. Allman and others reconsidered the idea and framed it as a social hypothesis (Allman 1999). Also known as the *Machiavellian intelligence hypothesis*, it was more rigorously formulated by R. Dunbar (2004) and extended into other mental and biomedical fields (Baron-Cohen et al. 1999; Badcok and Crespi 2008). Although the hypothesis has been criticized from several grounds (Balter 2012), and it is unclear whether it can be extended to the generality of mammalian societies, it has gained momentum regarding the evolutionary explanation of the 'natural' groups and structures formed in human societies. In the present work, the social brain views have been taken as one of the main references to structurally develop the sociotype hypothesis.

*1.2. Further relational and mental health aspects*

Nevertheless, the main argument of this paper will depart from the social brain hypothesis in two important respects. First, the emphasis will be put, not just in the size structures of social groups, but mostly in the communication practices that underlie the formation and maintenance of the individual's bonding networks --the relational, linguistic activities. Every interpersonal bond is but a "shared memory", consisting in specialized neural "engrams" that encode a variable number of *ad hoc* behavioral episodes positively or negatively finalized (Collins and Marijuán 1997). Being far more than collections of mere recognition events, bond memories would occupy an important quota of cortical space, presumably with each bond's occupancy depending on its 'strength'. This bonding reliance on vast cortical spaces would be in accordance with the relevant multi-area activations produced by social interactions and social evaluations, as observed in different neuroimaging studies (Greene 2001; Lacoboni 2004; Cacciopo and Patrick 2008). Hence, the overall cortical conformation and capacity of our species would greatly influence the number of bonds that, in general, human individuals can meaningfully sustain. However, like many other brain/mind phenomena, exactly how bonds are made, maintained, eroded, finalized, or restored is not sufficiently understood yet. Language appears to be the essential tool for bond making in human societies, although not the only one (Benzon 2001; Marijuán and Navarro 2010). Distinguishing several classes of bonds (related to their strength) would also be important in order to assess their respective relevance within the relational sociotype of the individual, and how the daily conversation/communication budget apportioned among the different bonding classes becomes sufficiently rewarding or not for each individual. Analyzing the different conversation-time distributions could lead to very interesting comparatives: by age, gender, status, professions, cultures…

The second aspect in which the present work departs from the social brain hypothesis concerns its empirical, or better, pragmatic orientation. Herein the emphasis will be put on elaborating a mental-health oriented construct, roughly exploring the potential applications of the sociotype as an indicator gauging the whole relational networks of the person, and how much daily conversation/communication he or she is engaged on a regular basis. Seemingly, rather than the exchange of functional information, it is trivial conversation, gossiping about social acquaintances what represents the human equivalent of primate grooming --subsequently stimulating in our "social brain" the production of endorphins, which relieve stress and boost the immune system (Dunbar 2004; Nelson 2007; Shutt et al. 2007). Thus, counting with an appropriate network of relationships that can provide us pieces of amusing conversation would be an essential ingredient to our social, psychological and physical well-being. Notwithstanding a number of recent studies on social networks (technologically oriented) that have tracked vast amounts of interpersonal exchanges, the metrics of the relational structures necessary for mental health and well-being have not been properly addressed yet. The hope is that the progressive delineation of a sociotype concept, pragmatically oriented, and susceptible of both theoretical and empirical demarcation, could contribute to a better understanding of the structures and dynamics of human sociality, and even provide some practical help when sociality itself is in crisis, as seem to be happening with the current "epidemics of loneliness" affecting large population tracts (Hawkley and Cacioppo 2010).

*1.3. Loneliness and its psychobiological consequences*

In our times the absence of social bonds has become a common experience: over 80% of children and 40% of those over 65 report feeling alone from time to time. Loneliness levels gradually decline in the middle years of adulthood and increase with age (reaching the maximum around age 70) (Weeks 1994; Pinquart and Sorensen 2001; Berguno et al. 2004). As numerous studies have shown, there is an association between social isolation, primarily perceived isolation, and poor physical and mental health, which cannot be explained away using different health behaviors. Social isolation decreases life years of social species, from *Drosophila* (Ruan and Wu 2008) to *Homo sapiens* (House et al. 1988). The lack of social bonds has deleterious effects on health through its effect on the brain, the hypothalamic-pituitary-adrenal (HPA), vascular processes, blood pressure, gene transcription, inflammatory, immune, and sleep quality (Cacioppo and Hawkley 2009). Research indicates that perceived social isolation (i.e., loneliness) is a risk factor, and may contribute to poorer cognitive performance, greater cognitive impairment and poorer executive function and an increased negativity and depressive cognition that accentuate sensitivity to social threats (Berkman 2009). In fact, loneliness is associated not only with poor physical health; it also includes psychiatric conditions such as schizophrenia and personality disorders, suicidal thoughts, depression and Alzheimer (Berkman 2009; Wilson et al. 2007; Cacioppo and Hawkley 2009).

*1.4. A growing social problem*

It's supposed we are living in a society "technologically civilized", where the ubiquitous presence of Media and Information Technologies has dramatically altered life styles. But it is unclear the effect that such pervasiveness of Media and Information Technologies and their overuse are having in our social relationships and quality of life. In what extent could computers, cell phones, and TVs replace our need of face-to-face relationships? Are they facilitators or surrogates and false substitutes? In today's society there is a significant change in the way social relationships are maintained, for the intrusion of the new ITs adds to the important social disintegration that is occurring for other reasons (aging, migration, marginalization of minorities, etc.). In our times, relational networks are apparently larger and faster, but more transient and devoid of personal contact, so that individuals are at greater risk of social isolation. The evidence in fast-developing countries is that economic growth and technological development have gone hand-in-hand with an increase in mental and behavioral disorders, family disintegration, social exclusion, and lower social trust (Bok 2010; Huppert 2010).

In 1950, 4 million Americans lived alone, making up 9% of households; the census data from 2011 show that nearly 33 million Americans are living alone, making up 28% of American households: three hundred per cent increase. The same process is taking place in different countries, for example in Sweden the percentage of households "single" reaches 47%, Britain 34%, 31% in Japan, 29% in Italy and 25% in Russia. Living alone, paradoxically, could symbolize our social need to reconnect (Klinenberg 2012). Similarly, mental disorders such as schizophrenia, depression, epilepsy, dementia, alcoholism and other substances abuse constitute 13% of the global disease burden, a percentage that surpasses cardiovascular diseases and cancer (Collins et al. 2011). European studies estimate that in the period of one year, 165 million people (38% of the population) will develop a mental illness (Wittchen et al. 2011). In Spain, according to the Time Use Survey (INE, 2010), people spend less and less time to interact physically and face to face. Between 2003 and 2010 participation in social life and fun activities decreased, while the time spent with computers (social networks, information retrieval, computer games) substantially increased, from 17.3% of population in 2003 to 30% in 2011. Socializing and fun activities were performed by 57% of the population, while seven years earlier (2003) these activities were performed by 64.4%. In recent years there has been a significant transfer of social life and collective fun activities to individualized activities such as computer games, Internet, TV watching. In this regard, it is significant that in Spain and in other countries, suicide rates have increased dramatically in the last three decades. In the US Census, 1985, the average number of confidants was three; in the 2004 census the average was 2, but the most common figure was *zero* confidants for almost 25 % (Cacciopo and Patrick 2008).

In spite of the pervasive epidemics of loneliness and lack of meaningful relationships in contemporary societies, there is a dearth of adequate indicators gauging the social networking and relational activities of the individual. How much do we talk along the day? With whom? Face-to-face, by phone, via Internet? How often do we socialize? Do we exercise alone? This type of questions has to be properly addressed and integrated with the measurement of the social networks around the individual, and further correlated with well-being and mental health questionnaires. That's what the sociotype hypothesis aims.

## 2. THE SOCIOTYPE HYPOTHESIS

The term sociotype has already appeared in the literature, though very scantly. In psychology, it has been put into use by a Jungian oriented school, "socionics", meaning the specific profile attributed to some well-recognized professions: lawyer, policemen, firefighter, etc. (Jung 1971). In the biomedical area, E.M. Berry (2011) has recently proposed the sociotype as an integrative term covering internal and external factors for the management of chronic disease, implying the integration of bio-psycho-sociology with systems biology. Also, some other authors have already utilized the term within the triad genotype-phenotype-sociotype, implying the social-evolutionary meaning herein proposed (Marijuán 2006, 2009; del Moral and Navarro 2012).

The sociotype construct is an attempt to cover the social interactions (bonding structures and communication relationships) that are adaptively demanded by the 'social brain' of each individual. In the same way that there is scientific consensus on the validity of the genotype and phenotype constructs for the human species, notwithstanding their respective degrees of variability, a metrics could also be developed applying to the relative constancy of the social environment to which the individuals of our species are evolutionarily adapted. The average brain stimulation coming from relational interactions in that social environment, together with further substitutes and surrogates culturally elaborated, would constitute a mental necessity for the individual's well being. Thus, the interest of appropriately gauging the bonding structures and communication relationships by means of a questionnaire, or a series of questionnaires, including also the influence of factors related to age, personality, culture, etc. This sociotype construct could provide relevant help for psychological counseling and early psychiatric intervention.

*2.1. Fundamental hypotheses*

More concretely, developing the sociotype construct would imply addressing and putting into test the following fundamental hypotheses:

- There exist in human beings an individual-specific and pervasive way to conform the structure and dynamics of individual social bonds, that probably can be related with the function of several brain structures.
- It is possible to develop a questionnaire to assess and measure this construct to validate it in the general population (quite probably in subpopulations segmented by age).
- The sociotype can be a useful indicator of mental and general health in the population, becoming an adjuvant tool for psychiatric diagnosis and risk assessment of mental illness.

Thereafter, the following objectives have been addressed:

*2.2. Central objective*

- The central goal is establishing a new indicator, based on a standard questionnaire, to collect essential data on the structure of the individual's social bonds, as well as their dynamic update (conversation), and correlate it with other indicators of mental health.

2.3. *Secondary objectives*

- Develop a questionnaire that can measure and validate the sociotype concept in the Spanish population.
- Generalize the sociotype concept and its indicator as a general means of social and psychological study.

- Framing the sociotype as an indicator of mental and general health in the various segments of the population (youth, adults, seniors).
- Demonstrate its use as adjuvant tool for the psychiatric diagnosis and risk for mental illness in patients with depression and dermatology patients.

*2.4. Pilot Study*

In order to address both the structural and dynamic aspects of the sociotype construct, a pilot study has been undertaken analyzing the social networks around adolescents. Subsequently, a sociotype questionnaire has been developed, initially tailored for the adolescent population. In this preliminary study we have selected a young population due to the high prevalence and intensity of the feelings of loneliness, actually higher in adolescence and transition to adulthood (16-25 years) than in any other group except the elderly (> 80 years) (Pinquart and Sorensen 2003). The study in older people has been discarded precisely because most research in loneliness has already been done in older population (Cacioppo and Hawkley 2009). A total of 165 students were interviewed with the preliminary version of the "Sociotype Test" developed by the authors. A group of 95 students was recruited from the first courses of University (19-20 years old), and another group of 70 students was recruited from the second and third courses of University (20-22 years old).

**3. METHODOLOGY**

*3.1. Design:* Exploratory, observational, cross-sectional study.

*3.2. Study population*

In this study we had applied a "convenience sampling" (Cohen et al. 2009), where the subjects were students who came from two education centers which we had access. The total sample of students was 165. There were two samples, *Sample_1* was recruited from last two courses of High School, its sample size was n=95 (38.3% men and 61.7% women) and the age average 19.37 (SD=2.44). *Sample_2* was composed by students from the first course of University, its sample size was n=70 (80% women and 20% men) and the average age was 21.42 (SD=1.37).

All the individuals were Spanish and none suffered any mental illness that prevented the realization of the task, so they were able to understand and complete the questionnaire. Inclusion criteria were as follows: age 18-25 years, good mastery of Spanish language. Exclusion criteria were: to suffer from severe mental disorder, any clinical o psychological illness that prevented the realization of the test.

Sample_1 was interviewed with the "Sociotype Test" in order to explore both structural and dynamic aspects of social networking, and Sample_2 was also interviewed with the GHQ-28 (General Health Questionnaire), addressed to relate the social networks with mental health and psychological well-being.

The Ethical Committee of Aragón had previously surveyed the Methodology of the Study, as part of the Project FIS PI12/01480. All subjects were students above 18 years, the questionnaires were anonymous, and the requested data didn't involve confidential aspects, so the informed consent was obtained verbally.

*3.3. Sociotype Questionnaire*

This preliminary Sociotype Questionnaire was developed by the authors. Based on the opinion of experts from different fields of knowledge (ej: sociology, anthropology, psychiatry/psychology, neuroscience) a set of 6-8 dominions to assess the concept were developed. Using qualitative methods (in-depth interviews, discussion groups, etc.), healthy people and patients with psychiatric and physical disorders were approached to identify the key questions to assess those dominions. Finally, factorial analysis was used to identify the definitive items included in the questionnaire following the usual methods to develop new questionnaires (Montero-Marín and García-Campayo 2010).

This preliminary version of the Sociotype Questionnaire was made up of 20 items. It included basic socio-demographic questions (age, gender, educational level, family status...), and also questions related to the way relationships are kept (time talking face-to-face, telephone, social networks or other channels). They were requested for the four different layers of social relationships considered (nuclear family, close friends, relatives and parenthood, social acquaintances). The auto-evaluation of sociability, as well as the self-satisfaction level was asked too; and also changes in personal state. The questionnaire showed adequate psychometric properties that will be described in an independent paper.

### *3.4. GHQ-28 Questionnaire*

The General Health Questionaire-28 is a screening tool to detect emotional distress and the risk of developing psychiatric disorders. Through factor analysis, the GHQ-28 considers four subscales: somatic symptoms, anxiety/insomnia, social dysfunction and severe depression.

The scoring method (CGHQ) takes into account the chronicity of psychiatric symptoms. It is superior to the conventional scoring method in yielding a wider range of scores, a more normal distribution and a well validated measure of neurotic illness. We used the validated Spanish version of the questionnaire (Lobo et al. 1986).

### *3.5. Statistical Analysis*

Frequency distributions of the qualitative variables were calculated in each category (gender, pets…). Quantitative variables (time talking, number of contacts…) were tested for normal distribution by means of Kolmogorov-Smirnov test, and indicators of central tendency (mean, trimmed mean or median) and dispersion (standard deviation or percentiles) were elaborated. Correlation between social variables and psychological risk factors were performed by means of contrast hypothesis, comparing proportions of qualitative variables (chi-square, Fisher exact test) or by comparison of means of quantitative variables (Student's t, ANOVA). When the distribution wasn't adjusted to normalcy, the U Mann Whitney or Kruskal Wallis tests were used. The analysis was executed by means of the SPSS 15.0 for Windows. A significance level (alpha) of 5% was used to consider statistical significance.

## 4. RESULTS

Structural and relational data were obtained. In the former, four levels of relationships were distinguished (arguably, three levels could have been better, as will be discussed in the next section). The results may be seen just in Table 1.

|  | Mean | SD |
|---|---|---|
| **Nuclear Family** | 5.05 | 1.24 |
| **Close Friends** | 6.02 | 3.23 |
| **Relatives&Parenthood** | 13.03 | 10.21 |
| **Social Acquaintances** | 77.06 | 92.85 |

**Table 1**. Number of people in the different layers of social relationships

About the relational data, they have been presented in minutes per week, for an easy calculation, and they are aggregated for the whole population (see Table 2). They are also compared by gender (Table 3), in that case including equivalence in hours per day.

|  | Mean | SD |
|---|---|---|
| **Family Face-to-Face** | 464.40 | 394.17 |
| **Family Phone** | 41.52 | 62.65 |
| **Couple Face-to-Face** | 495.19 | 342.39 |
| **Couple Phone** | 96.59 | 95.01 |
| **Friends Face-to-Face** | 492.33 | 380.50 |
| **Friend Phone** | 69.36 | 80.58 |
| **Acquaintances Face-to-Face** | 127.10 | 149.57 |
| **Acquaintance Phone** | 19.89 | 34.48 |

**Table 2.** Conversation time (min/weekly)

|  | Minutes per Week | | Hours per Day | |
|---|---|---|---|---|
|  | Women | Men | Women | Men |
| **Family Face-to-Face** | 473.89 | 421.11 | 1.13 | 1.00 |
| **Family Phone** | 58.98 | 31.85 | 0.14 | 0.08 |
| **Couple Face-to-Face** | 542.64 | 272.22 | 1.29 | 0.65 |
| **Couple Phone** | 101.34 | 35 | 0.24 | 0.08 |
| **Friends Face-to-Face** | 518.06 | 593.33 | 1.23 | 1.41 |
| **Friends Phone** | 108.06 | 29.44 | 0.26 | 0.07 |
| **Acquaintances Face-to-Face** | 131.81 | 131.11 | 0.31 | 0.31 |
| **Acquaintance Phone** | 15.83 | 12.96 | 0.04 | 0.03 |

**Table 3.** Time spent in conversation, by gender. 5% Trimmed mean

The statistical analysis of the relationship between the most relevant variables in the Sociotype Questionnaire and the General Health Questionaire-28 is shown in Table 4.

The data are corresponded to the four psychiatric subscales, and quite many of them show statistical significance.

|  | Somatic Symptoms | | Anxiety & Insomnia | | Social Dysfunction | | Severe Depression | |
|---|---|---|---|---|---|---|---|---|
|  | Not Probably | Probably | Not probably | Probably | Not probably | Probably | Not probably | Probably |
| **Gender (women)** | 33 (80.5%) | 23 (79.3%) | 21 (77.8%) | 35 (81.4%) | 54 (83.1%)* | 2 (40%)* | 50 (80.6%) | 6 (75%) |
| **Almost 3 people you can trust when facing a problem** | 39 (95.1%)* | 23 (79.3%)* | 26 (96.3%) | 36 (83.7%) | 59 (90.8%)* | 3 (60%)* | 57 (91.9%)* | 5 (62.5%)* |
| **People get in touch– weekly** | 53 (90) | 39 (27) | 45 (64)* | 41 (34)* | 42 (40) | 19 (36) | 44 (42.5) | 38 (32.5) |
| **Time talking – min/week** | 2760 (3555) | 1620 (4270) | 2880 (4620) | 2280 (4020) | 2430 (3965) | 1320 (3200) | 2385 (3927.5) | 1890 (4627.5) |
| **Tapas/Café – weekly** | 1 (2) | 0 (2) | 2 (3)* | 0 (2)* | 1 (2) | 0 (2) | 1 (2) | 0 (3.25) |
| **Go for a walk accompanied– weekly** | 3 (4) | 2 (4) | 4 (4.25)* | 2 (4)* | 3 (3)* | 0 (1.5)* | 2.5 (3) | 0.5 (3.5) |
| **Sociability Level** | 75 (12.5) | 80 (13.5) | 75 (15) | 80 (10) | 80 (10)* | 63 (45)* | 80 (10)* | 55 (39.5)* |
| **Satisfaction Personal Relationship** | 80 (20)* | 70 (25)* | 80 (20) | 80 (20) | 80 (15) | 80 (43.5) | 80 (15) | 73.5 (60) |

**Table 4.** Relations between the subscales psychiatrics and the variables of interest. Medians (Interquartile Range) are shown for the quantitative variables; Frequency (Percentage) for the qualitative variable. *indicates statistical significance ($p<0.05$)

## 5. DISCUSSION AND COMMENTS

To the authors' information, this is the first attempt to identify and measure a construct related with the way that human beings develop and maintain their structure of relational bonds. Neither the emerging structures nor the dynamic relationships have been studied in their mutual interaction yet. As mentioned in the Introduction, the very important changes in the patterns of socialization within the "information societies" demand more sophisticate conceptual apparatuses to better tackle the inherent problems.

Although in this discussion we have to refer to preliminary results obtained from a very limited survey, the topics which surface are of general interest and confirm the potential of the sociotype construct.

A number of observations can be made about the social structure depicted in Table 1. Overall, the whole data indicate a pattern of superimposed social structures with a consistent number of 100 members among young people (mean age=19.37, SD=2.44). Certainly this value is not much close to Dunbar's, but probably it is different due to the fact that the structural pattern at this age is not totally established. Actually the standard deviation for social acquaintances is even higher than the mean itself. A significant number of students responded with pretty low figures, while other referred to several hundreds. So to speak, at this age an independent sociotype is in the making, and for some adolescents that is an unwanted task, while for others an unbridled social excitation reigns. Correlating the type of structural values obtained with personality traits would be quite interesting –to be done in a near future. Besides, the number of layers or levels to distinguish is also an interesting aspect. In some societies, the 'extended family' layer makes little sense, while in others it becomes the fundamental strata (power of clans).

In view of the obtained results in Table 2 we confirm the average of 3-4 h of daily conversation time referred in the literature (Dunbar 2004). We find gender as a fundamental factor (Table 3): women spend 1 hour per day more than men in communication (4.64h women vs. 3.63h men). Obviously these data should be studied in more detail and with larger samples. Although the conversation time is not strictly correlated with any subscale, there is an evidence of higher probability of developing a psychiatric disorder when the talking time is too low. As we have already argued, conversational 'grooming' would be essential to our social, psychological and physical well-being.

Attending to the influence of social networks on mental health (see Table 4), we find very interesting correlations: *Severe Depression* is directly correlated with the number of people you can trust when facing a problem, as well as with the sociability level; the same parameters correlate to *Social Dysfunction*, plus gender and going for a walk accompanied, and gender; *Anxiety and Insomnia* also correlates with the number of people you get in touch weekly, going for a walk accompanied, and going out for 'tapas' or café weekly; *Somatic Symptoms* and the self-satisfaction with your personal relationships. These results emphasize that loneliness may be a risk factor by interfering with some forms of psychosocial distress. In this questionnaire the concept of loneliness is considered as a separate entity from social isolation and depression, so these measures of relationships include the distress that an individual may subjectively feel.

As a further step, theoretical network approaches could be applied to the present data. What is the equivalent in terms of structural sociotype of the gain and loss of bonds in the different categories? How the sociotype evolves with age? How resilient is this structure regarding changes in the social environment, e.g., migrations? How do contemporary technological-communication changes affect its dynamics? In what extent could computers, cell phones, and Internet exchanges accelerate our bonding relationships? May all those ITs gadgets replace our need of face-to-face contact? In what extent is continuous "accessibility" irrespective of the interpersonal environment a disturbing circumstance? This Pilot Study has shown an intriguing panorama of correlations to be explored carefully, hinting at comparative studies on age and cultural differences. It also conduces to highly debated topics on mental health and psychiatry, such as the therapeutic influence of changing life styles (Walsh 2011), which have to be urgently addressed by mental health professionals for fostering individual and social well-being, and for preserving and optimizing cognitive function. The social support concept and the so-called "buffering hypothesis" may also be considered under the sociotype angle (Qureshi et al. 2013).

The main limitations of the study are the following: First, the questionnaire used to assess the sociotype is preliminary and should be subject to more thorough validation; but the preliminary results suggest that the construct exists and can be measured. Second, sample size is relatively small and not representative of the general population. Future studies with larger populations and including both healthy people and patients with psychiatric and physical disorders are warranted. Third, the concept of sociotype should be related not only with psychological variables but also with more biological variables such as genetic/epigenetic, neuroimage or immuno-neuro-endocrine variables to confirm the validity of the construct.

In any case, the present work is but an exploratory attempt, and further research on the sociotype topic is under way in a national mental health project (Spanish FIS Project, Carlos III Health Institute).

## 6. EVOLUTIONARY CODA

From the evolutionary point of view, the present "epidemics of loneliness" is nonsense, an arbitrary imposition stemming from both socio-cultural and techno-economic automatisms that are scarcely understood in their self-generating complexity. Much of the burden on health systems, particularly in mental health, derives from social disintegration –the lack of a community in which people can talk and feel connected to each other. It has been proved that people with the most extensive social networks and the highest levels of social engagement have the lowest rates of physical and cognitive decline. But it is very difficult investigating levels of social engagement, and even more measuring them. In some occasions, coining a new scientific concept helps to advance more useful ideas and social policies. The sociotype hypothesis and questionnaire herein presented may somehow contribute.

**ANNEX 1**

**CUESTIONARIO DE SOCIOTIPO**

1. Edad ____

2. Sexo ☐ Femenino ☐ Masculino

3. Estado civil
   - ☐ Soltero con pareja estable
   - ☐ Soltero sin pareja estable
   - ☐ Casado
   - ☐ Separado
   - ☐ Viudo

4. Número de hijos ____

5. Nivel de estudios
   - ☐ Primarios
   - ☐ Bachiller/FP
   - ☐ Grado Medio
   - ☐ Estudios superiores

6. En el momento actual tu actividad principal es
   - ☐ Estudiante
   - ☐ Trabajador en activo
   - ☐ Desempleado
   - ☐ Jubilado
   - ☐ Baja laboral
   - ☐ Otros

7. Si te encuentras trabajando en este momento, ¿cómo es el tipo de trabajo que realizas?
   - ☐ Individual
   - ☐ En equipo
   - ☐ Cara al público

8. Número de personas con las que convives en casa
   - ☐ Ninguna, vivo sólo
   - ☐ Convivo con ____ personas más aparte de mí.

9. En caso de convivir con otras personas indica tu relación y cuántos conviven contigo. Convivo con:
   - ☐ Pareja (1 persona)
   - ☐ Padre y/o madre ( _ personas)
   - ☐ Hermano/s ( _ personas)
   - ☐ Hijos ( _ personas)
   - ☐ Amigos ( _ personas)
   - ☐ Compañeros ( _ personas)
   - ☐ Otros ( _ personas)

10. ¿Vives en casa con alguna mascota? ¿Cuántas?
    - ☐ No
    - ☐ Sí, tengo ___

11. Si realizas de forma individual algún tipo de actividad cultural (pintura, lectura, música...), indica cuántas horas a la semana:
    - ☐ 0-4 horas
    - ☐ 5-9 horas
    - ☐ 10 horas o más

12. Si tienes teléfono móvil ¿cuántos contactos tienes? ____
    ¿Con cuántos de ellos hablas a lo largo de la semana? ____

13. Si perteneces a alguna red social (facebook, tuenti, twitter...),
    ¿Cuántos contactos tienes? ____
    ¿Con cuántos de ellos tienes conversaciones de manera habitual? ____

14. A continuación me gustaría que pensaras en las personas que componen tu red social, es decir, personas que forman parte de tu vida de alguna forma, es decir, personas que de alguna manera pueden influir en ti, que has visto, con quienes has hablado o te has escrito últimamente.

    a. De tu familia directa (pareja, padres, hermanos, hijos) ¿cuántas personas incluirías? ___
    b. De otros familiares (primos, tíos, sobrinos...) ¿cuántas personas incluirías? ___
    c. De amigos cercanos ¿cuántas personas incluirías? ___
    d. De allegados y conocidos ¿cuántas personas incluirías? ___
    e. De las personas de tu trabajo o escuela, ¿cuántas personas incluirías? ___
    f. De las personas de clubes, organizaciones, asociaciones, gimnasios, grupos a los que perteneces, ¿cuántas personas incluirías? ___
    g. Número de otras personas con quien tienes contacto y no pertenecen a ninguna de las categorías anteriores que incluirías (vecinos, profesionales con quién tienes contacto y conoces su nombre...) ___

15. De estas personas que componen tu red social, ¿podrías estimar aproximadamente el tiempo total que has pasado con ellos o has hablado con ellos (por teléfono, internet) en la última semana?

    a. Pareja:     ___hora/s   ___minutos
    b. Familia:    ___hora/s   ___minutos
    c. Amigos:     ___hora/s   ___minutos
    d. Conocidos:  ___hora/s   ___minutos
    e. Personas de trabajo o escuela:    ___hora/s   ___minutos
    f. Personas de grupos, asociaciones: ___hora/s   ___minutos
    g. Otras personas de tu red no incluidas en categorías anteriores:
                                         ___hora/s   ___minutos

16. De las personas que conforman tu red social, ¿a cuántas acudirías ante un problema importante?

    - ☐ 0-2 personas
    - ☐ 3-5 personas
    - ☐ 6-8 personas
    - ☐ 9-11 personas
    - ☐ 12 o más personas

17. Durante la última semana, ¿has realizado alguna de las siguientes actividades de carácter social en las que te relacionaste con familiares, amigos o conocidos?

¿Cuántas veces en la última semana?

- ☐ He visitado o recibido visitas en casa (incluido comidas sociales, cafés..) _____
- ☐ He comido o cenado fuera de casa (con familiares o amigos) _____
- ☐ He ido de tapas, café, cañas, vinos en bares o cafeterías _____
- ☐ He salido de fiesta, copas, a bailar... _____
- ☐ He paseado en compañía _____
- ☐ Me he juntado para jugar a las cartas u otros juegos _____
- ☐ He hablado por teléfono con familiares y amigos _____
- ☐ He tenido correspondencia ordinaria (he escrito o leído cartas de familiares y amigos _____
- ☐ He tenido correspondencia electrónica (e-mail) con familiares y amigos _____
- ☐ He tenido comunicaciones por ordenador (chat, teléfono por Internet, videoconferencias) _____
- ☐ He asistido a algún grupo o club social, asociación, hogar del jubilado... _____
- ☐ Otras (especificar):
    a) _______________________________ _____
    b) _______________________________ _____

18. ¿Puedes señalar si ha habido cambios en el último año importantes que han supuesto un cambio en tus relaciones personales?

- ☐ Emparejamiento
- ☐ Ruptura de pareja
- ☐ Pérdida familiar
- ☐ Incorporaciones a la estructura familiar
- ☐ Pérdida o cambio de puesto de trabajo
- ☐ Cambio de amistades
- ☐ Cambio de ciudad o domicilio
- ☐ Otros (especificar)

19. A continuación por favor, valora del 0 a 100 tu grado de sociabilidad, según la siguiente escala:

0 ________________________________________ 100
Nada sociable                          Extremadamente sociable

*Grado sociabilidad: _____

20. Valora tu satisfacción personal con las relaciones sociales de 0 a 100, según la siguiente escala:

0 ________________________________________ 100
Nada satisfecho                         Totalmente satisfecho

*Satisfacción personal relaciones sociales: _____